\documentclass[10pt,twocolumn,journal]{IEEEtran}
\usepackage{latexsym}
\usepackage{float}
\usepackage{amsfonts}
\usepackage{amsbsy}
\usepackage{amssymb}
\usepackage{times}
\usepackage{graphicx}
\usepackage{enumerate}
\usepackage[usenames]{color}
\usepackage[dvips]{pstcol}
\usepackage{epstopdf}
\usepackage[caption=false]{subfig}
\usepackage{cite}
\usepackage{amssymb}
\usepackage{amsfonts}
\usepackage{graphicx}
\usepackage{epsfig}
\usepackage{psfrag}
\usepackage{xcolor}
\usepackage{amsfonts, bm}
\usepackage{hhline}
\usepackage{epstopdf}
\usepackage{cite}
\usepackage{color}
\usepackage{xcolor}
\usepackage{subfig}
\usepackage{verbatim}
\usepackage{multirow}
\usepackage{booktabs}
\usepackage{amsthm}
\usepackage{makecell}
\usepackage{units}
\usepackage{amsmath}

\usepackage{stfloats}

\IEEEoverridecommandlockouts

\usepackage[linesnumbered,ruled]{algorithm2e}

\definecolor{ccr}{RGB}{0,0,255}  
\definecolor{ccb}{RGB}{0,0,255}  
\usepackage{hyperref}
\hypersetup{hypertex=true,
	colorlinks=true,
	linkcolor=ccr,
	anchorcolor=ccr,
	citecolor=ccb}
\begin{document}
	\title{Voxel-CKM: Voxelized  Radio Frequency Radiance Fields for Fast and Few-Shot CKM Construction}
	\author{Hanlei Li, Guangyi Zhang, Kequan Zhou, Yunlong Cai, and Guanding Yu
		\thanks{
			
			H. Li, G. Zhang, K. Zhou,  Y. Cai,  and G. Yu are with the College of Information Science and Electronic Engineering, Zhejiang University, Hangzhou 310027, China (e-mail: hanleili@zju.edu.cn; zhangguangyi@zju.edu.cn; kqzhou@zju.edu.cn; ylcai@zju.edu.cn; yuguanding@zju.edu.cn).	
	}}
	
	\maketitle

	\begin{abstract}	
Channel knowledge maps (CKMs) are designed to predict channel state information (CSI) from  user locations, thereby enabling low-overhead CSI acquisition.
However, existing CKM construction methods often  require hours-to-days of training time and dense measurements, resulting in substantial deployment cost. In this paper, we propose Voxel-CKM, a novel voxelized radio frequency (RF) radiance field framework for  fast and few-shot CKM construction.
The core idea is to replace implicit neural representations with  explicit  voxel grids to efficiently capture the spatial variation of wireless channels.
Building upon this, we  further introduce a compact vector-matrix (VM) decomposition to parameterize these voxel grids using a small set of matrices and vectors, which significantly accelerates convergence and facilitates fast CKM construction.
To enable few-shot learning, we incorporate a transmitter prior as an inductive bias to guide the learning  process under sparse measurements. Additionally, a total-variation (TV) regularization loss is proposed to mitigate overfitting and stabilize
optimization. 
Experiments show that Voxel-CKM substantially accelerates training convergence and improves performance in the few-shot regime.
	\end{abstract}
	

\section{Introduction}
Accurate and timely acquisition of channel state information (CSI) is fundamental to reliable and high-rate wireless communications. However, as sixth-generation (6G) systems are expected to employ large-scale antenna arrays and ultra-wide bandwidths \cite{MIMO_1, MIMO_2,MIMO_3, MIMO_4,MIMO_5,MIMO_6,MIMO_7}, conventional pilot-based CSI acquisition incurs prohibitive overhead. To address this challenge,  the channel knowledge map (CKM) has emerged as a promising alternative \cite{CKM_1, CKM_2,CKM_3}.  By predicting CSI directly from user locations, CKM can substantially reduce the overhead of CSI acquisition.  Nevertheless, realizing this benefit hinges on the efficient construction of high-fidelity CKMs, which remains a critical challenge that warrants further investigation.

\subsection{Prior Work}
Existing approaches to CKM construction can be broadly classified into  two categories: model-driven and data-driven approaches. Model-driven approaches exploit domain knowledge of wireless propagation to construct CKMs with strong physical interpretability \cite{Krig, krig_2,pathloss_model,model_free_0, ray_trace,EM,CKM_ray_trace,CKM_Scatterer}. Specifically, \cite{ray_trace} developed a calibrated ray-tracing framework that incorporates diffuse scattering effects to improve the accuracy of CKM construction.  
In addition, a scatterer-based modeling approach was proposed in \cite{CKM_Scatterer}, where CKMs are constructed by parameterizing dominant scatterer responses and estimating them via iterative optimization. 
However, model-driven approaches rely heavily on modeling assumptions and are inherently sensitive to model mismatch, which may lead to performance degradation in complex propagation environments. 
In contrast, data-driven methods bypass explicit propagation modeling and instead learn the location-to-CSI  mapping directly from measurement data using deep learning techniques \cite{model_free_1, model_free_2, CKM_CNN, CKM_diff, CKM_diff_2, model_free_6 ,model_free_5, GAN_CKM}.
In particular, \cite{model_free_2} formulated CKM construction as an image inpainting problem and proposed a Laplacian pyramid-based framework for efficient channel reconstruction. Moreover, a diffusion-based generative model was introduced in \cite{model_free_6}, which captures channel structures via a conditional generation process, enabling improved modeling of propagation effects.  
In \cite{GAN_CKM}, CKM construction was cast as a style transfer problem and addressed using a conditional generative adversarial network (GAN), where a U-Net generator and a patch-based discriminator are jointly trained to generate high-fidelity CKMs.
Despite their promising performance, data-driven approaches rely heavily on large-scale training datasets and often exhibit limited generalization to unseen environments, thereby hindering  practical deployment.


More recently,  neural radiance field (NeRF) \cite{NeRF} and three-dimensional Gaussian splatting (3DGS) \cite{3DGS} techniques have  catalyzed a new line of research on CKM construction.
NeRF and 3DGS represent  3D scenes using  parameterized spatial units and synthesize scene views from arbitrary viewpoints via radiance field rendering, providing a physically grounded paradigm for modeling light propagation.
Leveraging the analogy between optical and radio frequency (RF) propagation, several studies have adapted these radiance field rendering techniques to wireless  communication scenarios \cite{NeRF2, NeWRF, F4_CKM, WRF_GS,gsrf, RF_3DGS}. 
As a pioneering effort, \cite{NeRF2} presented the NeRF$^2$ framework, which represents wireless  environments using a set of virtual radiation sources and derives channel responses via volumetric rendering.
Following a similar paradigm, a NeRF-inspired approach termed NeWRF was developed in \cite{NeWRF}, where direction-of-arrival (DoA) is incorporated as side information to enhance CSI inference performance.
In parallel, the authors of \cite{WRF_GS} proposed WRF-GS+, 
a framework that represents  wireless environments with deformable 3D Gaussian primitives, enabling efficient channel prediction.
In \cite{RF_3DGS},  a 3DGS-based scheme named RF-3DGS was proposed, which encodes multi-modal channel characteristics using spherical harmonics (SH) to facilitate accurate wireless channel synthesis.

\subsection{Motivation and Contributions}
Despite the progress made by existing methods, we can identify
two key limitations:
\begin{itemize}
	\item  
	\textit{Lack of construction efficiency:} In practical wireless systems, CKMs are expected to be  constructed efficiently to facilitate timely deployment and updates. However, existing CKM construction methods typically require hours-to-days of training time, indicating limited construction efficiency. This inefficiency mainly stems from complex network architectures and computationally intensive optimization procedures, which lead to high computational overhead and slow convergence. As a result, these methods struggle to support real-time adaptation, which hinders their practical deployment.
	
	
	
	\item \textit{Reliance on dense measurements:} Existing CKM construction methods generally require a large number of measurements to achieve satisfactory accuracy. However, this requirement is difficult to meet in practical systems, as collecting dense  measurements is costly and time-consuming. When only limited training samples are available, existing methods often struggle to capture the complex spatial variations of wireless channels, leading to significant performance degradation. This heavy reliance on dense measurements poses a major barrier to the practical deployment of CKM.
	
	
\end{itemize}

In this paper, we propose Voxel-CKM, a novel voxelized RF radiance field framework for  \textit{fast} and \textit{few-shot} CKM construction. Specifically, we  formulate a voxelized RF radiance field that efficiently captures the spatial variation of wireless channels. Unlike NeRF-style methods \cite{NeRF2,NeWRF} that represent the radiance field using computationally expensive multilayer perceptrons (MLPs), we adopt an explicit voxel grid representation, which allows continuous field queries via efficient trilinear interpolation, thereby substantially accelerating the learning process. Building upon the explicit voxel-grid representation, we  adopt a compact vector-matrix (VM) decomposition  to further enhance construction efficiency. By replacing dense voxel parameters with a small set of low-rank matrices and vectors, this decomposition reduces the number of trainable parameters and improves training stability,  enabling faster convergence and thereby mitigating the first limitation. 

To address the second limitation,  we propose that appropriate initialization plays a critical role in addressing sparse measurements, since initialization strongly influences early-stage optimization and  restricts the space of feasible solutions under few-shot conditions.
Guided by physical insight, we identify the transmitter as a natural anchor for initialization, owing to its dominant role in shaping the spatial variation of wireless channels. Accordingly,  a transmitter prior is  incorporated to bias the voxelized RF radiance field before training.
Concretely, we  instantiate the prior by initializing the voxel grid with radial Gaussian functions, assigning higher initial values to voxels near the transmitter and smoothly decaying the values with distance. This prior provides an inductive bias that guides optimization under sparse measurements, effectively improving sample efficiency. Furthermore, we incorporate a total-variation (TV) regularization \cite{TV} on the voxel grid to combat overfitting in limited-data scenarios. By penalizing abrupt variations between neighboring voxels, the regularization encourages local spatial coherence, improving robustness and generalization in the few-shot regime. Simulation results demonstrate that Voxel-CKM achieves  faster convergence than existing baselines and  exhibits great performance under challenging few-shot conditions. 

Our main contributions are summarized as follows:

\begin{itemize}
	\item We propose Voxel-CKM, a novel framework for fast and few-shot CKM construction.  To achieve this, we formulate a voxelized RF radiance field by replacing implicit neural representations with explicit voxel grids, which efficiently capture the spatial variation of wireless channels.
	
	\item We introduce a compact VM decomposition method to accelerate the optimization of voxelized RF radiance fields. By decomposing dense voxel parameters into a small set of matrices and vectors, the proposed method significantly reduces the number of trainable parameters, thereby improving optimization efficiency.
	
	\item We incorporate a transmitter prior to enable few-shot learning. This prior injects a  physically motivated inductive bias into the voxelized RF radiance field, guiding its optimization under sparse measurements. Additionally, a TV regularization term is proposed to  mitigate overfitting and improve robustness in the few-shot regime.
	
	\item We present extensive experimental results to validate the effectiveness of the proposed framework. Compared to existing methods, 
	Voxel-CKM achieves faster convergence during training and delivers superior performance under few-shot conditions.
\end{itemize}

\subsection{Organization and Notations}
The rest of  this paper is organized as follows. Section II introduces the preliminaries on wireless channel modeling and NeRF techniques. Section III formulates the voxelized RF radiance field. The construction of Voxel-CKM is detailed in Section IV. Then, simulation results are provided in Section V, and Section VI concludes the paper.

Unless otherwise specified, scalars, vectors, and tensors of dimension greater than one are denoted by lowercase, boldface lowercase, and boldface uppercase letters, respectively. For a scalar $a$, $|a|$
represents its absolute value. For  vectors and tensors, $\|\cdot\|_1$ and $\|\cdot\|_2$ denote the entry-wise $\ell_1$ norm and the Euclidean  norm, respectively.
Finally, 
${\mathbb{C}^{m \times n}}({\mathbb{R}^{m \times n}})$ is the space of ${m \times n}$ complex (real) matrices.

\section{Preliminaries}
In this section, we introduce the preliminaries of wireless
channel modeling and NeRF techniques.
\subsection{Wireless Channel Modeling}



A typical wireless communication system comprises a transmitter, a receiver, and a wireless propagation environment. The transmitter generates and modulates a signal  $x=A{e}^{j\varphi}$, where 
$A$ and $\varphi$ represent the amplitude and phase, respectively. 
 As the signal propagates through the environment,  it undergoes amplitude attenuation $\Delta A$ and phase rotation $\Delta \varphi$. According to the basic free space path loss model, the amplitude attenuation 
$\Delta A$ scales inversely with the propagation distance $d$, while the phase rotation $\Delta \varphi$ varies linearly with $d$. The received signal can therefore be written as
\begin{subequations}
	\begin{align}
		&y=x\cdot \Delta A e^{j\Delta  \varphi}
		=A{e}^{j\varphi}\cdot \Delta A e^{j\Delta \varphi}, \tag{1a} \\
		&\Delta A = \frac{c}{4\pi fd},\ \Delta \varphi = \frac{-2\pi fd }{c}, \tag{1b}
	\end{align}
\end{subequations}
where $f$ represents  the carrier frequency and $c$ denotes the speed of light. In realistic propagation environments, reflections and scattering cause the signal to propagate along multiple paths, each experiencing distinct attenuation and phase shifts.  Consequently, the received signal becomes the superposition of all multipath components, expressed as 
\begin{equation} \label{OFDM}
	y=A{e}^{j\varphi}\cdot \sum_{l=1}^{L}{\Delta A_l e^{j\Delta  \varphi_l}},
\end{equation}
where $L$ is the number of propagation paths,   $\Delta  A_l$
and $\Delta  \varphi_l$ represent the amplitude attenuation and phase rotation  of the $l$-th path, respectively.
The wireless channel $h$ is  defined as  
the ratio of the received signal to  the transmitted signal:
\begin{equation} \label{wireless_model}
	h \triangleq \frac{y}{x}=\sum_{l=1}^{L}{\Delta  A_l e^{j\Delta  \varphi_l}}.
\end{equation}

\subsection{NeRF Techniques}
NeRF   is a deep learning-based framework for 3D scene representation \cite{NeRF}.  It represents  a scene using a continuous function  parameterized by a neural network (NN). This function takes  as input a spatial coordinate $\mathbf{p}=(x,y,z)$ and a viewing direction $\bm{\omega}=(\theta,\phi)$, and outputs  the corresponding  volume density $\sigma$ and directional radiance $\mathbf{c}=(r,g,b)$. 


To render an image from a given viewpoint, NeRF casts rays  from each pixel and samples a set of points along  each ray.
For each sampled point,  its spatial coordinate and viewing direction are fed into the NN to obtain the  corresponding  volume density and radiance. The pixel color, $\hat{C}(\bm{\omega})$,  is then computed as a weighted sum of the radiance  values  at the sampled points:
\begin{equation}
	\hat{C}(\bm{\omega}) = 
	\sum_{j=1}^{N} T_j \alpha_j \mathbf{c}_j,
\end{equation}
where  $\alpha_j= 1 - e^{-\sigma_j \delta_j}$ and $T_j = \prod_{k=1}^{j-1} (1 -  \alpha_k)$ indicate the opacity and accumulated transmittance for the $j$-th sampled point, respectively,
both $j$ and $k$ denote the sample indices along the ray, $N$ is the number of sampled points, and $\delta_j = ||\mathbf{p}_{j+1} - \mathbf{p}_j||_2$ is the  distance between adjacent samples.  The synthesized image is then compared with the ground-truth image to compute a mean squared error (MSE) loss, which guides the optimization of the network parameters.


\section{Voxelized RF Radiance Field }
CKM construction  is often constrained by tight time budgets and sparse measurements, as environmental dynamics necessitate rapid updates while dense channel measurements are costly in deployment.
Motivated by these practical constraints, we address the problem of fast and few-shot CKM construction in a  wireless  communication scenario, where the transmitter is fixed at a known location $\mathbf{p}_{t}$.
In the few-shot setting, we assume that only $M$ sparse  measurements  $\{(\mathbf{p}_{r}^i, h^i)\}_{i=1}^{M}$  are available, where each $h^i \in \mathbb{C}$  denotes the CSI measured at receiver location $\mathbf{p}^i_{r}$. Our goal is to predict the CSI at arbitrary  receiver locations by efficiently learning a CKM  from these sparse measurements.

\begin{figure}[t]
	\centering 
	\includegraphics[width=1.0\linewidth]{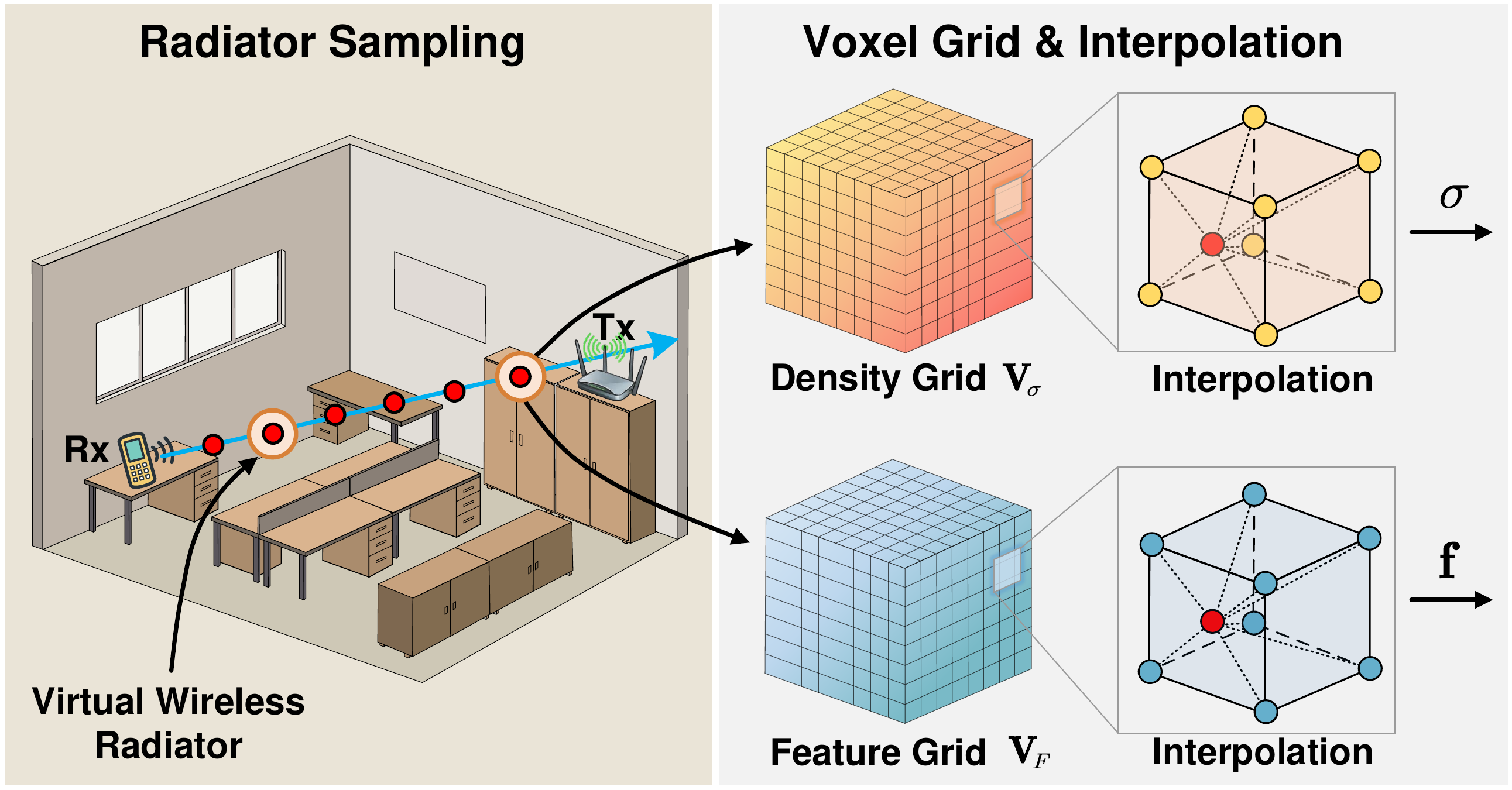} 
	\captionsetup{font={footnotesize}}
	\caption{Overview of the voxelized RF radiance field. The wireless environment is modeled as a set of virtual wireless radiators, whose radiance and density attributes are stored in explicit  voxel grids and queried via trilinear interpolation for CSI prediction.}
	\label{fig:voxelized_field} 
	\vspace{0pt}
\end{figure}



We begin by formulating the voxelized RF radiance field, as illustrated in Fig. \ref{fig:voxelized_field}.  The Huygens–Fresnel principle \cite{Huygens} states that any point reached by an RF signal can be regarded as a secondary source that re-radiates the signal outward. 
Motivated by this,  we treat each spatial point in the environment as a virtual radiator, each associated with two learnable attributes:
a complex radiance coefficient $s \in \mathbb{C}$ and a scalar density $\sigma \in \mathbb{R}$. Here, the complex coefficient $s$ represents the  directional radiance response, while the density $\sigma$ characterizes the local material absorption  of  RF signals.
Rather than implicitly encoding these radiance and density attributes using a deep MLP, we adopt an explicit voxel-based representation. Specifically, we construct a density grid $\mathbf{V}_{\sigma}\in \mathbb{R}^{I \times J \times K \times 1}$  and a feature grid  $\mathbf{V}_{F}\in \mathbb{R}^{I \times J \times K \times D}$, where  $I$, $J$, and $K$ denote the grid resolutions along the $X$, $Y$, and $Z$ axes, respectively, and $D$ represents the dimensionality of the  feature vector stored at each voxel in the feature grid. 

\begin{figure*}[t]
	\centering 
	\includegraphics[width=1.0\linewidth]{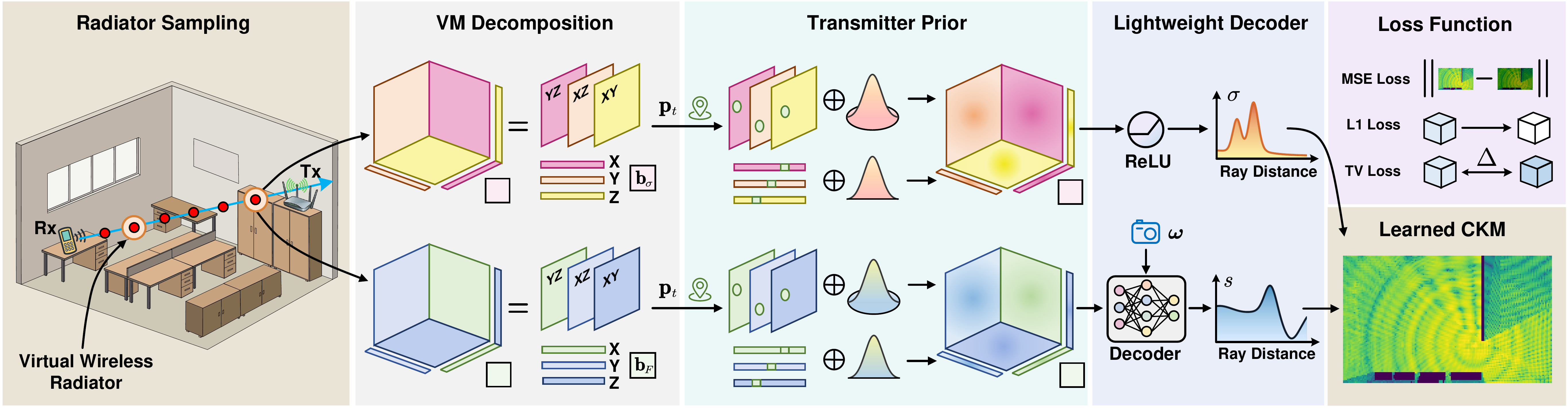} 
	\captionsetup{font={footnotesize  }}
	\caption{Pipeline of the Voxel-CKM framework.
		The framework builds on a voxelized RF radiance field and integrates a VM decomposition, a transmitter-aware prior, a lightweight decoder, and a regularized loss to support efficient CKM construction from sparse measurements.}
	\label{fig:Voxel_CKM} 
\end{figure*}

For an arbitrary 3D spatial point $\mathbf{p}=(x,y,z)$, its density value $\sigma(\mathbf{p})$ is obtained via trilinear interpolation over the density grid $\mathbf{V}_\sigma$, as shown in Fig. \ref{fig:voxelized_field}. Specifically, $\mathbf{p}$ is first mapped to its enclosing voxel cell, whose eight vertices are indexed in the density grid $\mathbf{V}_\sigma$. The density $\sigma(\mathbf{p})$ is then computed as a weighted combination of the density values stored at these eight vertices, where the weights are determined by the relative offsets of $\mathbf{p}$ within the voxel cell. This process is formally expressed as
\begin{equation} \label{calculation_for_sigma}
	\sigma(\mathbf p)
	=
	\sum_{n=1}^{8}
	w_n(\mathbf p)\mathbf V_\sigma[\mathbf q_n],
\end{equation}
where $\mathbf q_n$ denotes the $n$-th vertex of the voxel
cell enclosing $\mathbf p$, $\mathbf V_\sigma[\mathbf q_n]$ represents the density value
stored at vertex $\mathbf q_n$, and $w_n(\mathbf p)$ is the trilinear
interpolation weight.

Similarly, the radiance coefficient at the query point is obtained from the feature grid. We first query the feature grid $\mathbf{V}_{{F}}$  via trilinear interpolation to produce a feature vector $\mathbf{f}(\mathbf{p}) \in \mathbb{R}^D$. Since the interpolated feature $\mathbf{f}(\mathbf{p})$ is inherently agnostic of the viewing direction, we further  combine it with the radiance direction $\boldsymbol{\omega}$ and feed them into a lightweight decoder to predict the directional radiance coefficient $s(\mathbf{p}, \boldsymbol{\omega})$:
\begin{equation} \label{calculation_for_s}
	s(\mathbf{p}, \boldsymbol{\omega}) = \mathcal{D}_{\bm{\theta}}\!\left(\mathbf{f}(\mathbf{p}), \boldsymbol{\omega}\right),
\end{equation}
where $\mathcal{D}_{\bm{\theta}}(\cdot)$ denotes the lightweight decoder, and $\bm{\theta}$ represents its parameter set.

To predict the CSI at the receiver, we first define a set of rays $\{\mathbf{r}_i\}_{i=1}^{N_r}$, each originating from the receiver location and extending into the scene along its DoA, where $N_r$ denotes the number of rays.
The DoA can be obtained using the ray-searching algorithm in \cite{NeWRF}.
Along each ray $\mathbf{r}_i$, we  sample a set of discrete points $\{\mathbf{p}_{ij}\}_{j=1}^{N_s}$, where $i$  indicates the ray index and $N_s$ represents the number of samples per ray. 
By querying the voxel grids, the density $\sigma_{ij}$ and  radiance coefficient $s_{ij}$  at each sampled point are computed according to Eqs. (\ref{calculation_for_sigma}) and (\ref{calculation_for_s}).
Following standard volume rendering, the density values are converted into an absorption ratio $\alpha_{ij}$ and an accumulated transmittance $T_{ij}$:
\begin{equation} \label{calculation_for_T_and_alpha}
	T_{ij} = \prod_{k=1}^{j-1}(1 - \alpha_{ik}), \quad
	\alpha_{ij} = 1 - \exp \left( -\sigma_{ij} \delta_{ij}\right),
\end{equation}
where  $i$ denotes the ray index, both $j$ and $k$ represent the
indices of sampled points,  and  $\delta_{ij} = \|\mathbf{p}_{i,j+1} - \mathbf{p}_{ij}\|_2$ denotes the  distance between adjacent samples.
In addition to material absorption, RF signals also  experience  free-space path loss due to spherical wave spreading \cite{wave}. To model this attenuation, we introduce a free-space attenuation factor $\rho$, expressed as
\begin{equation} \label{calculation_for_rho}
	\rho_{ij}=\frac{c}{4\pi fd_{ij}} e^{ -j2\pi f d_{ij}/c},
\end{equation}
where $d_{ij}$ denotes the distance  between 
$\mathbf{p}_{ij}$ 	 and the receiver,  and
$f$	represents the  carrier frequency.
By aggregating the contributions  from all sampled points across all rays, the CSI prediction $\hat{h}$ at the receiver is computed as	
\begin{equation} \label{calculation_for_CSI}
	\hat{h} = 
	\sum_{i=1}^{N_r}\sum_{j=1}^{N_s} T_{ij}  \rho_{ij}  \alpha_{ij}  s_{ij}.
\end{equation}
The predicted CSI is then  compared with the ground-truth CSI to compute the training loss, which is used to jointly optimize the voxel grids  and the lightweight decoder.

\section{Construction of Voxel-CKM}
In this section, we provide an in-depth elaboration on the
proposed Voxel-CKM and its implementation specifics.

\subsection{Construction Pipeline}
   Building upon the voxelized RF radiance field, we present our
   Voxel-CKM framework, which integrates structured factorization with physics-informed priors for fast and few-shot CKM construction. The overall construction pipeline is illustrated in Fig. \ref{fig:Voxel_CKM}. In particular, Voxel-CKM consists of the following key components:
   
   \begin{itemize}
   	\item \textbf{VM Decomposition:} 
	This method parameterizes the voxel grids with compact vector and matrix factors. It preserves the efficiency of the explicit grid representation while significantly reducing the number of trainable parameters, thereby enabling faster convergence and  rapid adaptation.
   	\item \textbf{Transmitter Prior:} 
   	This prior initializes the vector and matrix factors with radial Gaussian functions centered at the transmitter location. 
   	The resulting initialization  encourages higher voxel values near the transmitter and a smooth decay with distance,    thereby  guiding optimization under sparse measurements.
   	\item \textbf{Lightweight Decoder:} 
   	This module predicts the radiance coefficient $s$  from the queried feature $\mathbf{f}(\mathbf{p})$ and the radiance direction $\bm{\omega}$. 
   	Instead of using deep MLPs as in NeRF-style methods, we combine SH  encoding with compact linear layers, yielding a lightweight decoder.
   	\item \textbf{Regularized Loss Function:} 
   	This  loss function combines the CSI prediction loss with $\ell_1$ and TV regularization  terms imposed on the VM factors. These regularization terms constrain excessive factor magnitudes and  suppress abrupt spatial variations, which stabilizes optimization in the few-shot regime.
   	\item \textbf{Progressive Training Strategy:} 
   	This strategy optimizes the vector and matrix factors in a coarse-to-fine manner. 
   	The field is first learned at a low spatial resolution to capture the global propagation structure, and is then progressively upsampled to recover finer spatial details.
   \end{itemize}
   
 The details of these components are presented in the following subsections.

\subsection{VM Decomposition}
While explicit voxel grids provide an intuitive and spatially structured representation of RF radiance fields, directly optimizing dense grid parameters can be inefficient \cite{TensorRF}. To improve parameter efficiency and accelerate convergence, we adopt a compact VM decomposition to parameterize the voxel grids.

The core idea of VM decomposition is to represent a dense 3D tensor using a compact set of plane matrices and line vectors.
Formally, a dense 3D tensor $\mathbf{T} \in \mathbb{R}^{I \times J \times K}$ can be parameterized as
\begin{equation}
	\mathbf{T}
	=
	\sum_{r=1}^{R}
	\Big(
	\mathbf{v}_{r}^{X} \circ \mathbf{M}_{r}^{YZ}
	+
	\mathbf{v}_{r}^{Y} \circ \mathbf{M}_{r}^{XZ}
	+
	\mathbf{v}_{r}^{Z} \circ \mathbf{M}_{r}^{XY}
	\Big),
\end{equation}
where $\mathbf{v}_{r}^{X} \in \mathbb{R}^I$, $\mathbf{v}_{r}^{Y} \in \mathbb{R}^J, \mathbf{v}_{r}^{Z} \in \mathbb{R}^K$ are learnable vector factors along  the three coordinate axes, 
$\mathbf{M}_{r}^{YZ} \in \mathbb{R}^{J \times K}, \mathbf{M}_{r}^{XZ} \in \mathbb{R}^{I \times K}, \mathbf{M}_{r}^{XY} \in \mathbb{R}^{I \times J}$ represent the  learnable matrix factors, $R$ denotes the decomposition rank, and $\circ$ represents the outer product.


In our voxelized RF radiance field, the voxel grids are represented as 4D tensors, with an additional channel dimension corresponding to voxel-wise attributes. We therefore extend the above VM decomposition to the  4D setting.
To handle the channel dimension while maintaining a compact parameterization, we associate each vector–matrix pair in the VM decomposition with a learnable coefficient vector:
\begin{subequations} \label{eq:vm-outer}
	\begin{align}
		\mathbf{V}_{\sigma}
		= \sum_{r=1}^{R_1} \Big(
		&\mathbf{v}_{\sigma,r}^{X} \circ \mathbf{M}_{\sigma,r}^{YZ} \circ \mathbf{b}^1_{\sigma,r} 
		+ \mathbf{v}_{\sigma,r}^{Y} \circ \mathbf{M}_{\sigma,r}^{XZ} \circ \mathbf{b}^2_{\sigma,r} \notag \\
		+ &\mathbf{v}_{\sigma,r}^{Z} \circ \mathbf{M}_{\sigma,r}^{XY} \circ \mathbf{b}^3_{\sigma,r} \Big), \label{eq:vm-sigma} \\
		\mathbf{V}_{F}
		= \sum_{r=1}^{R_2} \Big(
		&\mathbf{v}_{F,r}^{X} \circ \mathbf{M}_{F,r}^{YZ} \circ \mathbf{b}^1_{F,r} 
		+ \mathbf{v}_{F,r}^{Y} \circ \mathbf{M}_{F,r}^{XZ} \circ \mathbf{b}^2_{F,r} \notag \\
		+ &\mathbf{v}_{F,r}^{Z} \circ \mathbf{M}_{F,r}^{XY} \circ \mathbf{b}^3_{F,r} \Big), \label{eq:vm-F}
	\end{align}
\end{subequations}
where $R_1$ and $R_2$ represent the decomposition ranks for the density grid and feature grid, respectively, and  $\mathbf{b}^k_{\sigma,r} \in \mathbb{R}$ and $\mathbf{b}^k_{{F},r} \in \mathbb{R}^D$ for $k = 1, 2, 3$ denote learnable coefficient vectors associated with the $r$-th vector–matrix pair.
Once the voxel grids are parameterized by these vector and matrix factors, we no longer  explicitly reconstruct the dense 3D tensors or apply trilinear interpolation. Instead, the field is queried directly from the factorized representation. 
Given a 3D query point $\mathbf{p} = (x, y, z)$, we  project it onto the three coordinate axes ($X, Y, Z$) and the three canonical planes ($XY, XZ, YZ$). 
At these projected locations, we perform 1D linear interpolation on the corresponding vector factors and 2D bilinear interpolation on the corresponding matrix factors, yielding the line and plane coefficients\footnote{Here, vector and matrix factors refer to the learnable 1D and 2D arrays in the VM decomposition, respectively, whereas line and plane coefficients refer to the scalar
	values interpolated from these factors at the query point.}, respectively.
The density $\sigma(\mathbf{p})$ is then computed by combining these interpolated coefficients:
\begin{equation}\label{sigma_p}
	\begin{aligned}
		\sigma(\mathbf{p})\!=\!\sum_{r=1}^{R_1}\!\Big(
		&\mathbf{v}_{\sigma,r}^{X}(x)\,\mathbf{M}_{\sigma,r}^{YZ}(y,\!z)\,\mathbf{b}^{1}_{\sigma,r}
		\!+\!\mathbf{v}_{\sigma,r}^{Y}(y)\,\mathbf{M}_{\sigma,r}^{XZ}(x,\!z)\,\mathbf{b}^{2}_{\sigma,r}\\
		+
		&\mathbf{v}_{\sigma,r}^{Z}(z)\,\mathbf{M}_{\sigma,r}^{XY}(x,\!y)\,\mathbf{b}^{3}_{\sigma,r}
		\Big),
	\end{aligned}
\end{equation}
where $\mathbf{v}_{\sigma,r}^{X}(x)$, $\mathbf{v}_{\sigma,r}^{Y}(y)$, and $\mathbf{v}_{\sigma,r}^{Z}(z)$ denote the interpolated line coefficients evaluated at the projected locations, and $\mathbf{M}_{\sigma,r}^{XY}(x,y)$, $\mathbf{M}_{\sigma,r}^{XZ}(x,z)$, and $\mathbf{M}_{\sigma,r}^{YZ}(y,z)$ denote the  interpolated plane coefficients.
Using the same factorization and querying scheme, the feature  $\mathbf{f}(\mathbf{p})$ is computed as
\begin{equation} \label{f_p}
	\begin{aligned}
		\mathbf{f}(\mathbf{p})\!=\!\sum_{r=1}^{R_2}\!\Big(
		&\mathbf{v}_{F,r}^{X}(x)\,\mathbf{M}_{F,r}^{YZ}(y,\!z)\,\mathbf{b}^{1}_{F,r}
		\!+\! \mathbf{v}_{F,r}^{Y}(y)\,\mathbf{M}_{F,r}^{XZ}(x,\!z)\,\mathbf{b}^{2}_{F,r} \\
		+ &\mathbf{v}_{F,r}^{Z}(z)\,\mathbf{M}_{F,r}^{XY}(x,\!y)\,\mathbf{b}^{3}_{F,r}\Big).
	\end{aligned}
\end{equation}
The queried feature $\mathbf{f}(\mathbf{p})$ is then combined with the radiance direction $\bm{\omega}$ to predict the directional radiance coefficient.

\begin{figure}[t]
	\centering 
	\includegraphics[width=1.0\linewidth]{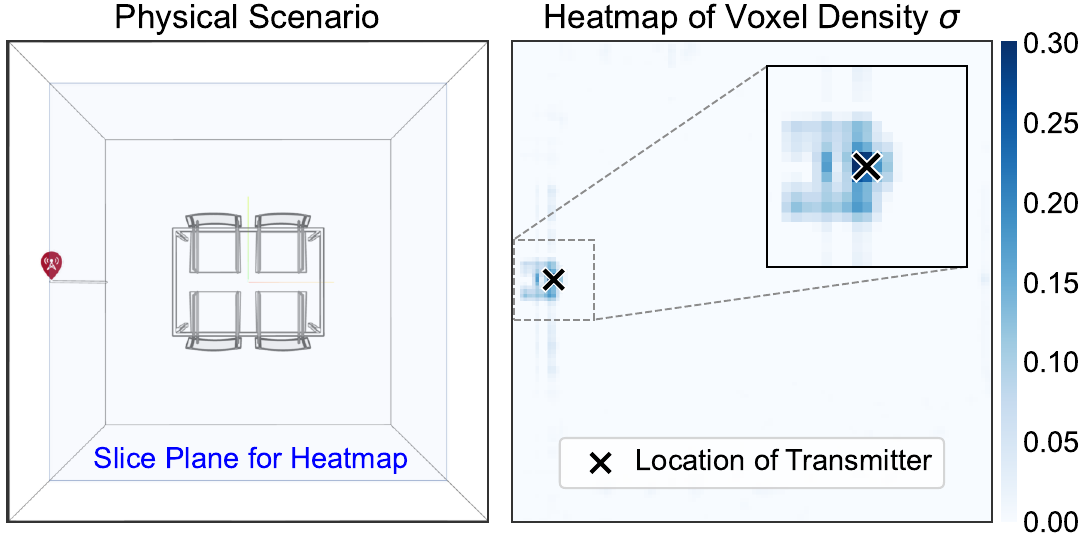} 
	\captionsetup{font={footnotesize  }}
	\caption{Visualization of the learned voxel density field. A slice plane is extracted to generate the heatmap of voxel density, which highlights the spatial concentration of voxel density around the transmitter location.}
	\label{fig:Transmitter_Prior} 
\end{figure}

\subsection{Transmitter Prior} 
In wireless propagation, the transmitter plays a dominant role in shaping the spatial variation of wireless channels, as the signal  originates  from the transmitter's location and its energy  gradually decays with distance.
This physical characteristic is also reflected in the learned radiance field, where higher values are observed near the transmitter, as illustrated in Fig. \ref{fig:Transmitter_Prior}.
Motivated by this  observation, we propose to explicitly incorporate a transmitter prior into the voxelized RF radiance field to guide  its optimization under sparse measurements. Concretely, the transmitter prior is introduced 
by initializing the  vector and matrix factors in the VM decomposition with radial Gaussian functions. The procedure is summarized as follows. For clarity, we focus on the vector factor $\mathbf{v}_{{F},r}^{X}$ and the matrix factor $\mathbf{M}_{{F},r}^{YZ}$, while the remaining VM factors follow the same procedure.


 \textit{1) Projecting Transmitter Location:} 
Given the transmitter coordinate $\mathbf{p}_{t} = (x_{t}, y_{t}, z_{t})$ and the scene bounds $[x_{\min}, x_{\max}] \times [y_{\min}, y_{\max}] \times [z_{\min}, z_{\max}]$, we first map the transmitter location  onto discrete  grid indices:
\begin{subequations} \label{prior_1}
	\begin{align}
		i_t &= \left\lfloor (x_t - x_{\min})(I-1)/(x_{\max}-x_{\min}) \right\rfloor, \label{prior_1_i} \\
		j_t &= \left\lfloor (y_t - y_{\min})(J-1)/(y_{\max}-y_{\min}) \right\rfloor, \label{prior_1_j} \\
		k_t &= \left\lfloor (z_t - z_{\min})(K-1)/(z_{\max}-z_{\min}) \right\rfloor, \label{prior_1_k}
	\end{align}
\end{subequations}
where  $i_t$, $j_t$, and $k_t$ denote the   indices of the transmitter along the $X$, $Y$, and $Z$ axes, respectively. Then, $i_t$ is used to index the entry of the $X$ axis vector factor $\mathbf{v}_{{F},r}^{X}$ and $(j_t, k_t)$ is used to index the entry of the $YZ$ matrix factor $\mathbf{M}_{{F},r}^{YZ}$.

 \textit{2) Generating Gaussian Biases:} 
According to the projected indices, we generate biases for the vector factor $\mathbf{v}_{{F},r}^{X}$ and the matrix factor $\mathbf{M}_{{F},r}^{YZ}$, respectively.
Specifically, these biases are  constructed using radial Gaussian functions centered at the projected  transmitter indices:
\begin{subequations} \label{prior_2}
	\begin{align}
		\mathbf{g}^{X}_{{F},r}[i]
		&= \eta_{v} \exp\!\left(-\frac{(i-i_t)^2}{2\sigma_v^{2}}\right), \label{prior_2_v} \\
		\mathbf{G}_{{F},r}^{YZ}[j,k]
		&= \eta_{m} \exp\!\left(-\frac{(j-j_t)^2+(k-k_t)^2}{2\sigma_{m}^{2}}\right), \label{prior_2_M}
	\end{align}
\end{subequations}
where $\eta_{v}$ and $\eta_{m}$ control the strengths of the biases, respectively, and $\sigma_{v}$ and $\sigma_{m}$ determine their spatial spread.






 \textit{3) Injecting Biases into VM Factors:} The generated biases are then added to the corresponding VM factors as initialization offsets   before training:
\begin{equation} \label{prior_3}
	\mathbf{v}_{{F},r}^{X} \leftarrow \mathbf{v}_{{F},r}^{X}+\mathbf{g}^{X}_{{F},r}, \quad
	\mathbf{M}_{{F},r}^{YZ} \leftarrow \mathbf{M}_{{F},r}^{YZ}+\mathbf{G}_{{F},r}^{YZ}.
\end{equation}
This initialization biases the   voxelized RF radiance field  toward a transmitter-centered spatial  structure, encouraging higher values near the transmitter and a smooth decay with distance, thereby  facilitating optimization under sparse measurements.

\subsection{Lightweight Decoder Design}
\begin{figure}[t]
		\centering 
		\includegraphics[width=0.9\linewidth]{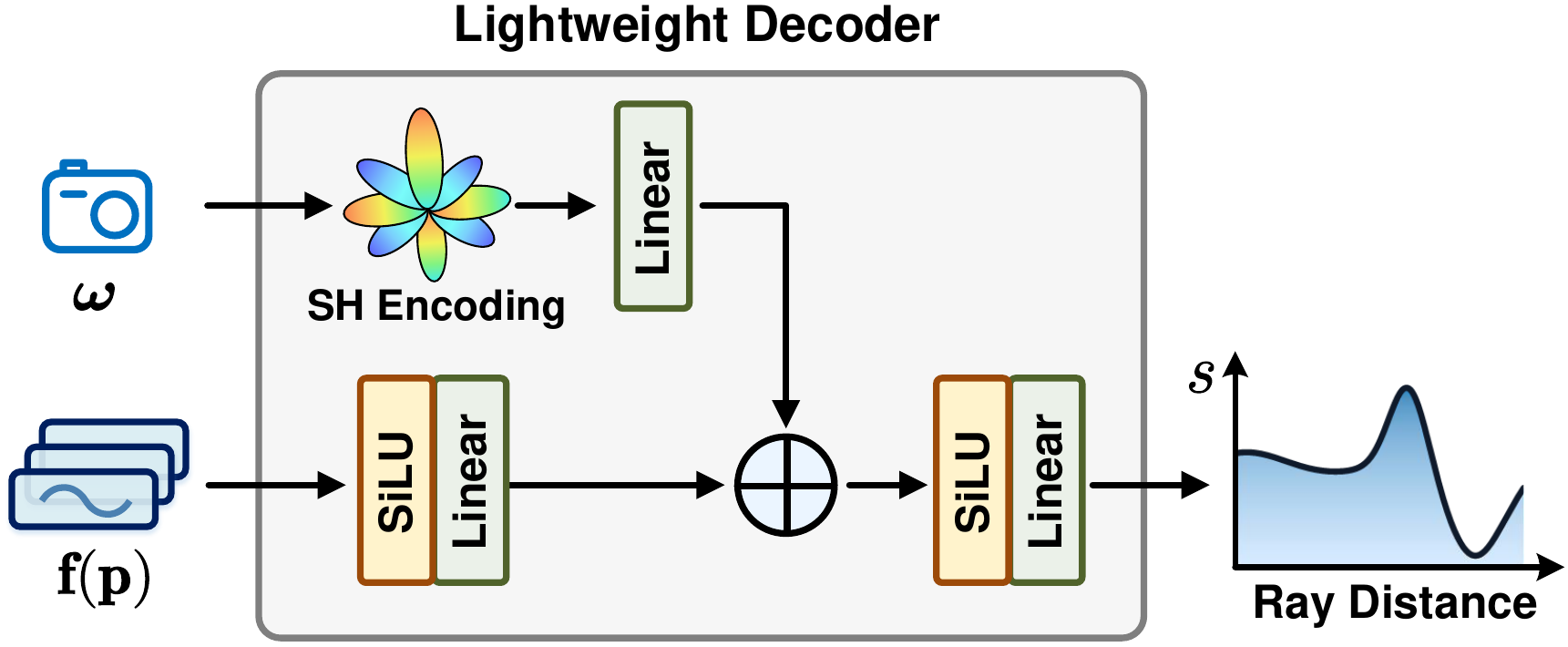} 
		\captionsetup{font={footnotesize  }}
		\caption{Architecture of the  proposed lightweight decoder.}
		\label{fig:deocder} 
	\end{figure}

Given the interpolated feature $\mathbf{f}(\mathbf{p})$  from the VM representation and the radiance direction $\bm{\omega}$, we employ a lightweight decoder to predict the directional  radiance coefficient $s(\mathbf{p}, \bm{\omega})$ of the virtual radiator.
As shown in Fig. \ref{fig:deocder}, the  decoder follows a two-branch architecture.
The first branch passes the interpolated feature $\mathbf{f}(\mathbf{p})$ through a SiLU activation \cite{Silu} and a linear layer, yielding a compact feature representation. The second branch first encodes the radiance direction $\bm{\omega}$ via SH encoding, which is defined as
\begin{equation}
\gamma_{\text{SH}}(\boldsymbol{\omega})
=
\{Y_\ell^m(\boldsymbol{\omega})\}_{0 \leq \ell \leq L_{\text{SH}},\,-\ell \leq m \leq \ell},
\end{equation}
where $Y_\ell^m$ denotes the SH basis function of degree $\ell$ and order $m$, and $L_{\text{SH}}$ is the maximum SH degree used in the encoding. The encoded direction is then mapped to a feature representation using a linear layer.
The outputs of the two branches are fused  through element-wise addition, followed by a SiLU activation  and a final linear layer to produce the radiance coefficients. These design choices keep the decoder lightweight by using a fixed SH basis for directional encoding and compact linear projections for feature transformation, rather than relying on the stacked hidden layers commonly adopted in NeRF-style MLP decoders.


	
	\subsection{Regularized  Loss Function}
	We optimize Voxel-CKM using a composite objective that combines a CSI prediction loss with  regularization terms on the VM factors.
The overall loss  function is defined as
\begin{equation} \label{loss_function}
	\mathcal{L}=\mathcal{L}_{\text{MSE}}+\lambda_{\ell_1}\mathcal{L}_{\ell_1}+\lambda_{\text{TV}}\mathcal{L}_{\text{TV}},
\end{equation}
where $\lambda_{\ell_1}$ and $\lambda_{\text{TV}}$ denote the hyperparameters to control the strength of the regularization terms.
The MSE term is defined as 
\begin{equation}
\mathcal{L}_{\text{MSE}} = \frac{1}{B}\sum_{i=1}^B \big| h^i - \hat{h}^i \big|^2,
\end{equation}
where $B$ represents the batch size. This term penalizes the discrepancy between  the predicted CSI $\hat{h}^i$ and the ground-truth CSI $h^i$. The $\ell_1$ term suppresses excessive magnitudes in the VM factors to stabilize optimization, expressed as
\begin{equation}
	\mathcal{L}_{\ell_1}
	=
	\sum_{\mathbf{v} \in \mathcal{V}}
	\frac{1}{N_{\mathbf{v}}} \|\mathbf{v}\|_1
	+
	\sum_{\mathbf{M} \in \mathcal{M}}
	\frac{1}{N_{\mathbf{M}}} \|\mathbf{M}\|_1,
\end{equation}
where $\mathcal{V}$ and $\mathcal{M}$ denote the sets of vector and matrix factors in the VM decomposition, respectively, and $N_{\mathbf{v}}$ and $N_{\mathbf{M}}$ represent the numbers of elements in $\mathbf{v}$ and $\mathbf{M}$, respectively.
To mitigate overfitting under sparse measurements, we further introduce a TV regularization term. This regularization penalizes the squared differences between adjacent VM elements, defined as
\begin{equation}
	\mathcal{L}_{\text{TV}}= 	
	\sum_{\mathbf{v} \in \mathcal{V}}
	\frac{1}{N_{\mathbf{v}}}
	\|\Delta \mathbf{v}\|^2_2
	+
	\sum_{\mathbf{M} \in \mathcal{M}}
	\frac{\beta}{N_{ \mathbf{ M}}}
	\|\Delta \mathbf{M}\|^2_2,
\end{equation}
where $\beta$ controls the relative contributions of vector and matrix factors, and $\Delta \mathbf{v}$ and $\Delta \mathbf{M}$ denote the differences between adjacent elements in the vector and matrix factors, respectively.  
Concretely, for a vector factor $\mathbf{v}\in\mathbb{R}^{I}$ and a matrix factor $\mathbf{M}\in\mathbb{R}^{J\times K}$, $\|\Delta\mathbf{v}\|^2_2$ and $\|\Delta\mathbf{M}\|^2_2$ are computed as
\begin{subequations} \label{eq:tv-detail}
	\begin{align}
		\|\Delta\mathbf{v}\|^2_2
		=& \sum_{i=1}^{I-1}\big(\mathbf{v}[i+1]-\mathbf{v}[i]\big)^2, \label{eq:tv-v} \\[0.5em]
		\|\Delta\mathbf{M}\|^2_2
		=& \sum_{j=1}^{J-1}\sum_{k=1}^{K}
		\big(\mathbf{M}[j+1,k]-\mathbf{M}[j,k]\big)^2 \notag \\
		\quad +
		& \sum_{j=1}^{J}\sum_{k=1}^{K-1}
		\big(\mathbf{M}[j,k+1]-\mathbf{M}[j,k]\big)^2. \label{eq:tv-M}
	\end{align}
\end{subequations}
By discouraging high-frequency variations in the VM factors, the TV regularization suppresses non-physical oscillations that may arise under sparse measurements, facilitating few-shot learning.

\begin{algorithm}[t] \label{alg:train}
	\fontsize{9}{10.9}\selectfont
	\DontPrintSemicolon
	\SetAlgoLined
	\KwIn{Training dataset $\mathcal{S}$; number of iterations $Q$; upsampling schedule $\mathcal{U}$.}
	\KwOut{Learned VM factors for $(\mathbf{V}_{\sigma},\mathbf{V}_{F})$ and lightweight decoder parameters.}
	Initialize VM factors and lightweight decoder.\;
	Apply transmitter prior to VM factors based on Eqs.~(\ref{prior_1}), (\ref{prior_2}), and~(\ref{prior_3}).\;
	\For{$q \gets 1$ \KwTo $Q$}{
		Sample a receiver location $\mathbf{p}_{r}$ from $\mathcal{S}$.\;
		Generate ray set $\{\mathbf{r}_{i}\}_{i=1}^{N_r}$ and sampled points $\{\mathbf{p}_{ij}\}_{j=1}^{N_s}$ along each ray.\;
		Query VM-factorized field to obtain density $\sigma_{ij}$ and radiance coefficient $s_{ij}$ for each sampled point based on Eqs.~(\ref{sigma_p}) and~(\ref{f_p}).\;
		Compute accumulated transmittance $T_{ij}$ and free-space attenuation factor $\rho_{ij}$ using Eqs.~(\ref{calculation_for_T_and_alpha}) and~(\ref{calculation_for_rho}).\;
		Compute the predicted CSI $\hat{h}$ using Eq.~(\ref{calculation_for_CSI}).\;
		Compute the training loss $\mathcal{L}$ based on Eq.~(\ref{loss_function}).\;
		Update the parameters of VM factors and lightweight decoder.\;
		\If{$q \in \mathcal{U}$}{
			Upsample vector factors by 1D linear interpolation and matrix factors by 2D bilinear interpolation.\;
		}
	}
	\caption{Training algorithm for Voxel-CKM}
\end{algorithm}

\subsection{Progressive Training Strategy}
Directly optimizing the VM factors at the target spatial resolution is inefficient,  since the high-resolution factors introduce excessive degrees of freedom, making the optimization susceptible to poor local minima during the early stage of training \cite{TensorRF}. 
To address this challenge, we adopt a progressive training strategy.
Specifically, this strategy initializes the VM factors at a coarse resolution and progressively increases their resolution at predefined training iterations. The intermediate resolutions are selected in logarithmic space between the initial and target resolutions, allowing the representation capacity to increase gradually throughout training. To facilitate a smooth transition across resolutions, the vector factors are upsampled along their axis via 1D linear interpolation, and the matrix factors are upsampled along their two axes via 2D bilinear interpolation. 
This coarse-to-fine strategy allows the model to capture global structure before refining details, thereby improving optimization efficiency. 

Algorithm \ref{alg:train}  summarizes the training procedure of Voxel-CKM. The VM factors and the lightweight decoder are first initialized, after which the transmitter prior is injected into the VM factors. At each iteration, we sample a receiver location, cast rays along its DoAs, and sample discrete points along each ray. The VM-factorized field is then queried at the sampled points to obtain density values and radiance coefficients, which are aggregated via differentiable rendering  to predict the CSI. The resulting prediction is used to compute the loss defined in Eq. (\ref{loss_function}), and the VM factors and decoder parameters are jointly optimized via gradient-based updates. Following a predefined schedule, the VM factors are progressively upsampled during training, enabling  efficient coarse-to-fine optimization.

\section{Experiments}
\subsection{Experimental Settings}

\textit{1) Implementation Details:} 
We implement  Voxel-CKM using the PyTorch platform and  conduct  experiments on a $2.60$ GHz Intel(R) Xeon(R) Platinum 8350C CPU and an NVIDIA GeForce RTX 4090 GPU.
All models are trained with a batch size of $32$ for $30,000$ iterations using the Adam optimizer. The learning rate is set to $2 \times 10^{-2}$ for VM factors and $1 \times 10^{-3}$ for the lightweight decoder. We apply an exponential learning rate decay, reducing the learning rate to $0.1$ of its initial value by the end of optimization. To balance optimization stability and model flexibility across different sampling budgets, we fix $\lambda_{\ell_1}=8\times10^{-5}$ across all experiments, while decreasing the TV regularization strength $\lambda_{\text{TV}}$ and the coefficient $\beta$ as the number of training samples $M$ increases. Specifically, $(\lambda_{\text{TV}}, \beta)$ is  set to $(10,1)$, $(1,0.1)$, and $(0.01,0)$ for $M=100$, $400$, and $1,000$, respectively.
These settings are shared across all scenes under the same sampling budgets.

\renewcommand{\arraystretch}{1.48}
\begin{table}[t]
	\centering
	\caption{Training settings and hyperparameters for the radiance field.}
	\resizebox{1.0\linewidth}{!}{
		\begin{tabular}{@{}ll|ll@{}}
			\toprule
			\multicolumn{2}{c|}{\textbf{Training Setting}} 
			& \multicolumn{2}{c}{\textbf{Hyperparameters for Radiance Field}} \\
			\midrule
			Platform & PyTorch & Samples per Ray $N_s$ & $128$ \\
			Optimizer & Adam & Density Rank $R_1$ & $16$ \\
			Batch Size & $32$ & Feature Rank $R_2$ & $48$ \\
			Iterations & $30{,}000$ & Feature Dim. $D$ & $27$ \\
			VM Factors LR  & $2 \times 10^{-2}$ & Bias Strength $\eta_v,\eta_m$ & $1$ \\
			Decoder LR & $1 \times 10^{-3}$ & Bias Spread $\sigma_v,\sigma_m$ & $3$ \\
			LR Decay & Exponential & Hidden Dim. $N_{\text{hid}}$ & $128$ \\
			$\ell_1$  Strength $\lambda_{\ell_1}$ & $8\times10^{-5}$ & Initial Resolution & $128\!\times\!128\!\times\!128$ \\
			TV  Strength $\lambda_{\text{TV}}$ & $\{10,1,0.01\}$ & Target Resolution & $300\!\times\!300\!\times\!300$ \\
			TV Coefficient $\beta$ & $\{1,0.1,0\}$ & Upsampling Schedule & $2\mathrm{k}$, $3\mathrm{k}$, $4\mathrm{k}$, $5\mathrm{k}$ \\
			\bottomrule
	\end{tabular}}
	\label{training_para}
\end{table}

\begin{figure}[t]
	\centering 
	\includegraphics[width=1.0\linewidth]{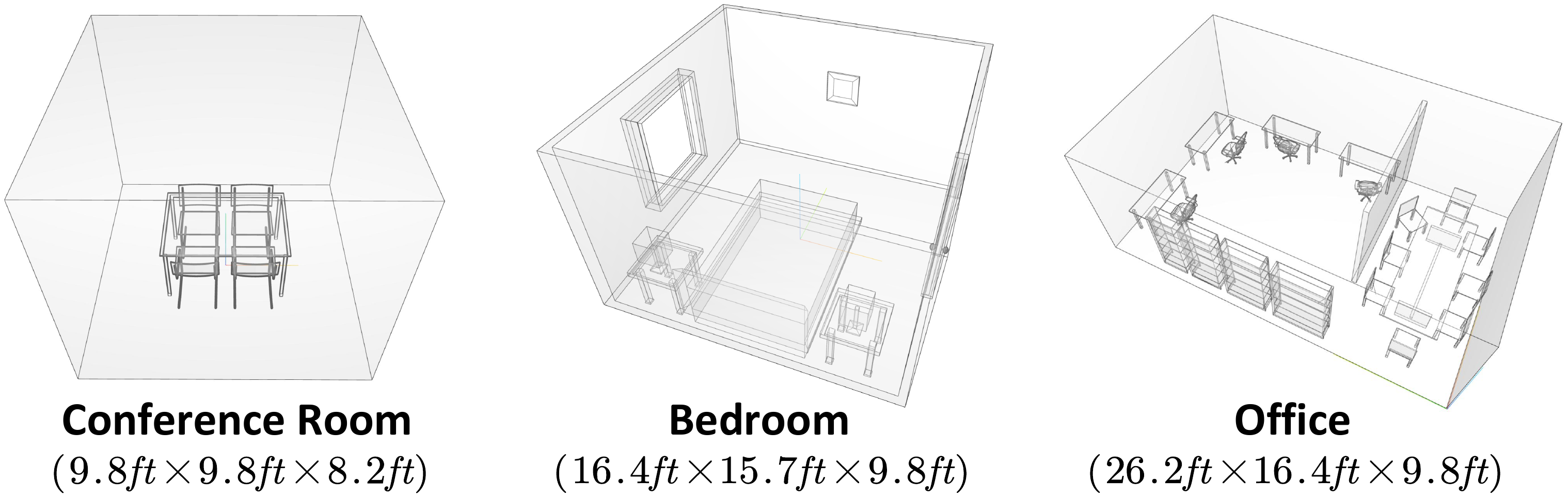} 
	\captionsetup{font={footnotesize  }}
	\caption{The geometries of three indoor 3D environment models.}
	\label{fig:geometry} 
\end{figure}

For the radiance field configuration, we uniformly sample $128$ points along each ray. 
The decomposition ranks for the density and feature grids are set to $R_1=16$ and $R_2=48$, respectively.  Each voxel in the feature grid stores a $27$-dimensional feature vector.  For the transmitter prior, the Gaussian bias strengths are set to $\eta_v=\eta_m=1$, and the bias spreads are set to $\sigma_v=\sigma_m=3$. The hidden dimension of the lightweight decoder is configured to $N_{\text{hid}}=128$. Under the progressive training strategy, the grid resolution is initialized at $128 \times 128 \times 128$ and gradually increased to $300 \times 300 \times 300$ during training. This upsampling is performed in logarithmic space at training iterations $2{,}000$, $3{,}000$, $4{,}000$, and $5{,}000$.  These parameter settings are summarized in Table~\ref{training_para}.

\textit{2) Datasets:} We evaluate the proposed method on both simulated and real-world datasets.
 \begin{itemize}
	\item  \textbf{Simulated Dataset:} 
In a manner similar to \cite{NeWRF}, we construct simulated datasets  across three indoor environments, including a conference room, a bedroom, and an office. The geometries of these environments are shown in Fig. \ref{fig:geometry}. In each scene, a single transmitter is placed at a fixed location, while receiver locations are randomly sampled within the environment. CSI samples are generated using the shooting and bouncing ray tracing method \cite{SBR} at a carrier frequency of $2.412$ GHz. For each scene, we generate $10,000$ training samples and $500$ test samples, where each sample corresponds to a distinct receiver location. 
For evaluation, we  subsample $100$, $400$, and $1,000$ training samples from the full training set, whereas the test set remains unchanged.
  	\item \textbf{Real-World Dataset:} 
In addition, we evaluate Voxel-CKM on a real-world dataset collected by Katholieke Universiteit Leuven~\cite{dataset}. 
The dataset contains $252{,}004$ over-the-air multiple-input multiple-output (MIMO) orthogonal frequency-division multiplexing (OFDM) channel measurements acquired in an indoor environment. The base station (BS) is equipped with $64$ antennas and the user equipment (UE) employs a single antenna.
The measurements are conducted using an OFDM system operating at a center frequency of $2.61$~GHz with $100$ subcarriers. 
To align with our problem formulation, the target CSI is defined as the channel coefficient corresponding to the first BS antenna and the first OFDM subcarrier.
For evaluation, we randomly sample $100$, $400$, and $1,000$ measurements to construct few-shot training sets, while an additional $10,000$ samples are used for testing.
\end{itemize}

\begin{table*}[t]
	\centering
	\caption{Few-shot performance comparison across three indoor environments.}
	\label{tab:few-shot}
	\setlength{\tabcolsep}{0pt}
	\renewcommand{\arraystretch}{1.45}
	\begin{tabular}{
			@{}
			>{\hspace{0.15cm}\raggedright\arraybackslash}m{2.5cm}|
			>{\centering\arraybackslash}m{1.65cm}
			>{\centering\arraybackslash}m{1.65cm}
			>{\centering\arraybackslash}m{1.65cm}|
			>{\centering\arraybackslash}m{1.65cm}
			>{\centering\arraybackslash}m{1.65cm}
			>{\centering\arraybackslash}m{1.65cm}|
			>{\centering\arraybackslash}m{1.65cm}
			>{\centering\arraybackslash}m{1.65cm}
			>{\centering\arraybackslash}m{1.65cm}
			@{}
		}
		\toprule
		\multirow{2}{*}{\raisebox{-2.8ex}{\textbf{CPSNR} (dB) $\uparrow$}}
		& \multicolumn{3}{c|}{\textbf{Conference Room}}
		& \multicolumn{3}{c|}{\textbf{Bedroom}}
		& \multicolumn{3}{c}{\textbf{Office}} \\
		& $M\!=\!100$ & $M\!=\!400$ & $M\!=\!1,000$
		& $M\!=\!100$ & $M\!=\!400$ & $M\!=\!1,000$
		& $M\!=\!100$ & $M\!=\!400$ & $M\!=\!1,000$ \\
		\midrule
		NeWRF \cite{NeWRF}         & $12.69$ & $21.08$ & $22.55$ & $6.55$ & $10.48$ & $12.08$ & $4.05$ & $7.89$ & $8.51$ \\
		$\text{NeRF}^2$ \cite{NeRF2} & $-1.48$ & $-1.22$ & $-1.37$ & $-1.53$ & $-2.38$ & $-1.90$ & $-1.92$ & $-1.45$ & $-1.64$ \\
		MLP \cite{MLP_new}         & $-1.66$ & $-1.71$ & $-1.94$ & $-2.11$ & $-1.88$ & $-1.82$ & $-2.21$ & $-1.83$ & $-2.15$ \\
		KNN \cite{KNN_new}         & $-0.76$ & $-1.05$ & $-0.93$ & $-1.16$ & $-1.30$ & $-0.38$ & $-1.45$ & $-1.26$ & $-1.01$ \\
		\midrule
		Voxel-CKM (Ours)           & $\mathbf{17.46}$ & $\mathbf{23.30}$ & $\mathbf{26.93}$ & $\mathbf{11.26}$ & $\mathbf{14.17}$ & $\mathbf{15.09}$ & $\mathbf{7.22}$ & $\mathbf{9.40}$ & $\mathbf{9.80}$ \\
		\bottomrule
	\end{tabular}
\end{table*}
\begin{figure*}[t]
	\centering 
	\includegraphics[width=1.0\linewidth]{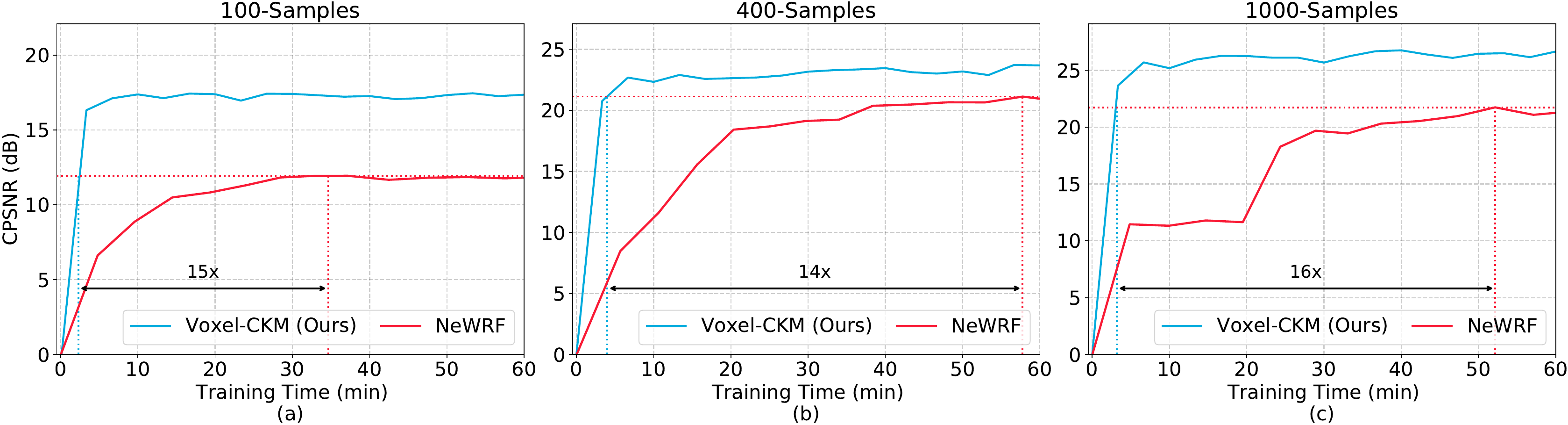} 
	\captionsetup{font={footnotesize  }}
	\caption{Comparison of training convergence curves between Voxel-CKM and NeWRF under different sampling budgets: (a) $100$ samples, (b) $400$ samples, (c) $1,000$ samples. We mark the time required for Voxel-CKM to attain the peak CPSNR of NeWRF and indicate the corresponding relative speedup.}
	\label{fig:Fast learning} 
\end{figure*}

\textit{3) Metrics and Baselines:} To evaluate the performance of Voxel-CKM, we employ the channel prediction signal-to-noise ratio (CPSNR),  defined as
\begin{equation} \label{CPSNR}
	\text{CPSNR}(\text{dB})=-10\log _{10}\frac{|\hat{h}-h|^2}{|h|^2},
\end{equation}
where $\hat{h}$ and $h$ denote the predicted and ground-truth CSI, respectively. A higher CPSNR implies more accurate  channel prediction performance.
 To validate the effectiveness of our model, we compare Voxel-CKM with several baselines.
\begin{itemize}
	 \item \textbf{NeWRF} \cite{NeWRF}: NeWRF is a NeRF-based framework that predicts  CSI directly from receiver locations. We adopt its open-source implementation and retrain the model on our  datasets for a fair comparison.
	\item \textbf{NeRF$^2$} \cite{NeRF2}: NeRF$^2$ is a channel modeling method originally designed for dynamic transmitter scenarios. We use its open-source code and adapt it to our setting.
	\item \textbf{MLP} \cite{MLP_new}: MLP learns a direct mapping from  receiver locations to CSI using a fully connected  network.
	\item \textbf{K-Nearest Neighbors (KNN)} \cite{KNN_new}: KNN predicts CSI at a test location by averaging the CSI values of its $K$ nearest neighbors in the training set, where  we set $K=3$ in all experiments.
\end{itemize}

\subsection{Evaluation  over Simulated Datasets}
\textit{1) Few-Shot Performance:} Table \ref{tab:few-shot} presents the few-shot  performance across three indoor environments under different sampling budgets. 
It is evident that our Voxel-CKM consistently outperforms all the baselines, with particularly pronounced advantages in the extreme few-shot regime.
When only $100$ training samples are available, most baseline methods suffer from severe performance degradation, exposing their strong reliance on dense measurements. In contrast, Voxel-CKM maintains stable and reliable prediction performance across all environments. This robustness can be attributed to the explicit voxelized representation and transmitter-aware prior, which provide strong inductive bias under sparse measurements.
As the training set grows to $400$ and $1,000$ samples, Voxel-CKM continues to deliver superior performance, achieving CPSNR improvements of up to $2$ dB over the strongest baseline.
These results underscore the reliability of Voxel-CKM, which remains effective across a wide range of sampling budgets.

\textit{2) Fast Learning Behavior:} To evaluate the fast-learning behavior of our approach, we  analyze the training convergence curves of Voxel-CKM under different sampling budgets in the conference room. 
As illustrated in Fig. \ref{fig:Fast learning}, Voxel-CKM exhibits faster convergence  than NeWRF across all sampling budgets,  requiring only about one-fifteenth of the training time to reach comparable performance.
This acceleration stems from the explicit voxelized RF  radiance field, which provides a structured representation that accelerates the learning process. As a result, Voxel-CKM  achieves highly efficient construction  and supports rapid  adaptation to new data, making it suitable for practical  scenarios.

\begin{table}[t]
	\centering
	\caption{Performance comparison on the real-world dataset under different sampling budgets $M$.}
	\label{tab:real_world}
	\renewcommand{\arraystretch}{1.6}
	\setlength{\tabcolsep}{2pt}   
	\begin{tabular}{
			>{\raggedright\arraybackslash}p{1.65cm}
			>{\raggedright\arraybackslash}p{1.85cm}
			|>{\centering\arraybackslash}p{1.5cm}
			>{\centering\arraybackslash}p{1.5cm}
			>{\centering\arraybackslash}p{1.5cm}
		}
		\toprule
		Method & Metric & $M\!=\!100$ & $M\!=\!400$ & $M\!=\!1{,}000$ \\
		\midrule
		\multirow{2}{*}{NeWRF \cite{NeWRF}}
		& CPSNR (dB) $\uparrow$ & $0.07$ & $2.98$ & $7.54$ \\
		& Time (min) $\downarrow$ & $98.55$ & $96.57$ & $99.08$ \\
		\midrule
		\multirow{2}{*}{$\text{NeRF}^2$ \cite{NeRF2}}
		& CPSNR (dB) $\uparrow$ & $-1.09$ & $-0.67$ & $-0.57$ \\
		& Time (min) $\downarrow$ & $160.88$ & $163.02$ & $167.99$ \\
		\midrule
		\multirow{2}{*}{Voxel-CKM}
		& CPSNR (dB) $\uparrow$ & $\mathbf{5.34}$ & $\mathbf{8.71}$ & $\mathbf{13.36}$ \\
		& Time (min) $\downarrow$ & $\mathbf{9.55}$ & $\mathbf{9.47}$ & $\mathbf{9.66}$ \\
		\bottomrule
	\end{tabular}
\end{table}

\subsection{Evaluation over Real-World Datasets}
To further demonstrate the practical applicability of our
approach, we compare Voxel-CKM with baseline methods
on a real-world MIMO dataset collected by Katholieke Universiteit Leuven \cite{dataset}. This dataset captures realistic propagation characteristics and reflects
practical non-idealities arising from hardware impairments, providing a challenging evaluation scenario.
Among the evaluated methods, NeWRF and Voxel-CKM require DoA information for ray sampling, which is not explicitly provided in the real-world dataset. To address this issue, we adopt a simple geometric approximation based on the known transmitter and receiver locations. Specifically, we first compute the line-of-sight (LoS) direction from the transmitter to the receiver, and then uniformly sample $18$ additional directions over the  angular domain, yielding a total of $19$ candidate DoAs. This DoA configuration is applied consistently to all DoA-dependent methods throughout the experiments.

Table~\ref{tab:real_world} summarizes the performance comparison on the real-world dataset under different sampling budgets.
Voxel-CKM delivers higher CPSNR than NeWRF and NeRF$^2$ across all settings.
The performance gap is particularly pronounced in the extremely sparse setting with $M=100$, where the baseline models obtain CPSNR values close to $0$~dB, suggesting limited prediction fidelity in the few-shot regime.
In contrast, Voxel-CKM remains effective and achieves a CPSNR of $5.34$~dB, demonstrating its stability and robustness under practical deployment scenarios.
In addition, Voxel-CKM requires substantially lower training time than NeWRF across all evaluated settings, indicating its efficiency  for practical applications.

\begin{table}[t]
	\centering
	\caption{Performance comparison under dense sampling in the conference room scenario.}
	\label{tab:dense-conference}
	\renewcommand{\arraystretch}{1.6}
	\setlength{\tabcolsep}{2pt}   
	\begin{tabular}{
			>{\raggedright\arraybackslash}p{1.65cm}
			>{\raggedright\arraybackslash}p{1.85cm}
			|>{\centering\arraybackslash}p{1.5cm}
			>{\centering\arraybackslash}p{1.5cm}
			>{\centering\arraybackslash}p{1.5cm}
		}
		\toprule
		Method & Metric & $M\!=\!3{,}000$ & $M\!=\!6{,}000$ & $M\!=\!10{,}000$ \\
		\midrule
		\multirow{2}{*}{NeWRF \cite{NeWRF}}
		& CPSNR (dB) $\uparrow$ & $23.85$ & $24.34$ & $24.61$ \\
		& Time (min) $\downarrow$ & $98.50$ & $96.44$ & $101.21$ \\
		\midrule
		\multirow{2}{*}{$\text{NeRF}^2$ \cite{NeRF2}}
		& CPSNR (dB) $\uparrow$ & $-1.39$ & $-0.88$ & $-0.27$ \\
		& Time (min) $\downarrow$ & $168.87$ & $170.10$ & $167.55$ \\
		\midrule
		\multirow{2}{*}{Voxel-CKM}
		& CPSNR (dB) $\uparrow$ & $\mathbf{27.83}$ & $\mathbf{29.05}$ & $\mathbf{29.64}$ \\
		& Time (min) $\downarrow$ & $\mathbf{17.40}$ & $\mathbf{17.48}$ & $\mathbf{17.58}$ \\
		\bottomrule
	\end{tabular}
\end{table}

\subsection{Evaluation under Extended Settings}
In this subsection, we conduct additional experiments to evaluate the robustness of Voxel-CKM under extended settings. Specifically, we consider two scenarios: 1) dense measurements and 2) the  OFDM setting.

\textit{1) Performance under Dense Measurements:}
In Table~\ref{tab:dense-conference}, we evaluate the performance of different methods under dense measurements  in the conference room scenario at $2.412$ GHz. Specifically, the number of training measurements is increased to 
$M=3,000$, $6,000$, and $10,000$.
To accommodate these larger training sets, we train our models for $50{,}000$ iterations, while keeping all other experimental settings unchanged.

As reported in Table~\ref{tab:dense-conference}, Voxel-CKM consistently achieves  higher CPSNR 
across all dense sampling configurations. Notably, the performance of Voxel-CKM continues to improve as the number of measurements increases, demonstrating its ability to effectively exploit additional supervision beyond the few-shot regime. 
Additionally, Voxel-CKM maintains significantly lower training time compared to the baseline methods. 
This behavior can be attributed to the explicit voxelized RF radiance field representation, which enables efficient CKM construction across a wide range of sampling budgets.

\textit{2) Performance under OFDM Setting:}
We further evaluate the proposed Voxel-CKM under an OFDM transmission setting to examine its ability to model frequency-selective channels.
Unlike the single-carrier case, an  OFDM system models  the wireless channel as a set of parallel subcarrier responses in the frequency domain, where the CSI is represented as a vector across subcarriers.
Following prior work \cite{NeWRF},  we construct a simulated OFDM dataset using the MATLAB platform. The OFDM system operates at a $20$ MHz bandwidth with $64$ subcarriers, among which $52$ subcarriers are used for channel measurements. All experiments are conducted in the conference room scenario, and results are reported for different measurement budgets $M$.

To extend Voxel-CKM to the OFDM setting, we introduce two straightforward modifications. First, the density grid is augmented with an additional subcarrier dimension, resulting in a grid of size 
$I \times J \times K \times N_c$, where $N_c=52$ represents the number of subcarriers. Second, the output dimension of the lightweight decoder is adjusted to jointly  predict the radiance coefficients across  all subcarriers.
For a fair comparison, the baseline methods are adapted to the OFDM setting by adjusting their output layers to match the dimension of the CSI vector. In the OFDM setting, CPSNR is computed over the multi-subcarrier CSI vector, defined as
\begin{equation}
	\text{CPSNR}(\text{dB})=-10\log _{10}\frac{||\mathbf{\hat{h}}-\mathbf{h}||^2_2}{||\mathbf{h}||^2_2},
\end{equation}    
where $\mathbf{\hat{h}}\in \mathbb{C}^{N_c}$ and $\mathbf{h}\in \mathbb{C}^{N_c}$ denote the predicted and ground-truth CSI vectors, respectively.

\begin{table}[t]
	\centering
	\caption{Performance comparison on the OFDM setting under different sampling budgets $M$ in the conference room scenario.}
	\label{tab:ofdm}
	\renewcommand{\arraystretch}{1.6}
	\setlength{\tabcolsep}{2pt}   
	\begin{tabular}{
			>{\raggedright\arraybackslash}p{1.65cm}
			>{\raggedright\arraybackslash}p{1.85cm}
			|>{\centering\arraybackslash}p{1.5cm}
			>{\centering\arraybackslash}p{1.5cm}
			>{\centering\arraybackslash}p{1.5cm}
		}
		\toprule
		Method & Metric & $M\!=\!100$ & $M\!=\!400$ & $M\!=\!1{,}000$ \\
		\midrule
		\multirow{2}{*}{NeWRF \cite{NeWRF}}
		& CPSNR (dB) $\uparrow$ & $12.34$ & $19.88$ & $25.19$ \\
		& Time (min) $\downarrow$ & $142.50$ & $145.55$ & $144.88$ \\
		\midrule
		\multirow{2}{*}{$\text{NeRF}^2$ \cite{NeRF2}}
		& CPSNR (dB) $\uparrow$ & $-1.71$ & $-1.55$ & $-1.47$ \\
		& Time (min) $\downarrow$ & $188.55$ & $189.05$ & $191.20$ \\
		\midrule
		\multirow{2}{*}{Voxel-CKM}
		& CPSNR (dB) $\uparrow$ & $\mathbf{15.66}$ & $\mathbf{20.08}$ & $\mathbf{26.79}$ \\
		& Time (min) $\downarrow$ & $\mathbf{28.50}$ & $\mathbf{28.95}$ & $\mathbf{29.17}$ \\
		\bottomrule
	\end{tabular}
\end{table}

As shown in Table~\ref{tab:ofdm}, Voxel-CKM outperforms existing methods across different measurement budgets.
This indicates  that the proposed Voxel-CKM can effectively capture  frequency-selective propagation characteristics  across multiple subcarriers, leading to accurate CSI prediction in OFDM settings.
In terms of training efficiency, although the CSI dimensionality increases substantially due to multi-subcarrier modeling, Voxel-CKM achieves  a training time of approximately $29$ minutes, while the other methods require more than $100$ minutes.
These results demonstrate that the proposed Voxel-CKM still supports fast learning even in challenging OFDM settings.

\subsection{Complexity Analysis}
To assess the efficiency and practicality of Voxel-CKM, we conduct a comprehensive complexity analysis from two aspects: space complexity and computational complexity. The space complexity quantifies the asymptotic storage cost of the learnable parameters, while the computational complexity measures the number of operations required for a forward pass. Let $N_{\text{SH}}$ denote the feature dimension after SH encoding. The detailed complexity analysis is presented as follows.

\textit{1) Space Complexity:} The space complexity of Voxel-CKM arises from the VM factors and the lightweight decoder.
\begin{itemize}
	\item \textbf{VM Factors:}
	Under the VM decomposition in Eq.~(\ref{eq:vm-sigma}), the density grid is represented by $R_1$ components, each consisting of three vector factors, three matrix factors, and three scalar coefficients. Ignoring constant factors, the corresponding storage cost is $O\big(R_1\big(I+J+K+IJ+IK+JK)\big)$. 	The feature grid follows the same VM structure with rank $R_2$, except that each scalar coefficient is replaced by a $D$-dimensional vector. This gives a complexity of
	$O\big(R_2(I+J+K+IJ+IK+JK+D)\big)$.
	Assuming $I$, $J$, and $K$ are $O(N)$, the overall space complexity of the VM factors can be given as
\begin{equation} 
	O\big((R_1+R_2)N^2+R_2D\big).
\end{equation} 
    Compared with dense voxel grids, which require $O(N^3D)$ parameters, the VM decomposition reduces the storage cost from cubic to quadratic order in the grid resolution.
	\item \textbf{Lightweight Decoder:}
	The lightweight decoder transforms the SH-encoded direction and interpolated feature  into a complex radiance coefficient. Ignoring biases and constant factors, it introduces a storage cost of
	\begin{equation} 
		O\big(N_{\text{SH}}N_{\text{hid}}+DN_{\text{hid}}\big),
	\end{equation} 
	where the two terms correspond to the linear projections applied to the SH-encoded  direction and the interpolated feature, respectively.
\end{itemize}
In summary, the space complexity of Voxel-CKM is
	\begin{equation} 
\mathcal{C}_{\text{space}}=
O\big((R_1+R_2)N^2+R_2D
+(N_{\text{SH}}+D)N_{\text{hid}}\big).
	\end{equation} 
This complexity can be controlled via the grid resolution $N$, the VM ranks $R_1$ and $R_2$, the feature dimension $D$, the SH encoding dimension $N_{\text{SH}}$, and the hidden dimension of the lightweight decoder $N_{\text{hid}}$.

\textit{2) Computational Complexity:}
The computational complexity of Voxel-CKM is dominated by three operations: field querying, lightweight decoding, and radiance rendering.
\begin{itemize}
	\item \textbf{Field Querying:}
	For each sampled point, the interpolation operation incurs an $O(1)$ cost for each VM factor. The per-point querying cost is therefore determined  by the subsequent aggregation steps.
	According to Eq.~(\ref{sigma_p}), the density value is computed by aggregating $R_1$ components, leading to a cost of $O(R_1)$. For the feature grid, the aggregation is performed over $R_2$ components across $D$ feature dimensions, which gives a cost of $O(R_2D)$. Therefore, the complexity of  per-point field querying is $O(R_1+R_2D)$.
	Since the query is evaluated at every sampled point along each ray, the overall complexity is
	\begin{equation} 
	O\big(N_rN_s(R_1+R_2D)\big).
	\end{equation} 
	
	\item \textbf{Lightweight Decoding:}
	The computational cost of the decoder is dominated by the linear projections of the SH-encoded direction and the interpolated feature. Considering $N_r$ rays and $N_s$ sampled points per ray, the computational complexity of the decoder is
	\begin{equation} 
	O\big(N_rN_sN_{\text{hid}}(N_{\text{SH}}+D)\big).
	\end{equation} 
	
	\item \textbf{Radiance Rendering:}
The computational complexity of radiance rendering mainly arises from computing the accumulated transmittance $T_{ij}$ and aggregating the radiance contributions according to Eq.~(\ref{calculation_for_CSI}). Since these operations involve only element-wise scalar computations, the resulting rendering complexity is
	\begin{equation} 
O\big(N_rN_s\big).
\end{equation}
\end{itemize}

Combining the costs of field querying, lightweight decoding, and radiance rendering, the overall computational complexity of Voxel-CKM is given by
	\begin{equation}
\mathcal{C}_{\text{comp}}
=
O\Big(
N_rN_s
\big[
R_1+R_2D+N_{\text{hid}}(N_{\text{SH}}+D)
\big]
\Big),
\end{equation}
which scales linearly with  the number of rays \(N_r\) and the per-ray radiator count \(N_s\). Notably, this complexity is independent of the grid resolution, since the forward pass queries the voxelized RF radiance field only at sampled points rather than traversing the entire grid. This resolution-independent query cost supports efficient CKM construction even when high-resolution grids are used.

\renewcommand{\arraystretch}{1.5}
\begin{table}[t]
	\centering
	\caption{Ablation study on the VM decomposition.}
	\label{tab:vm-ablation}
	\begin{tabular}{@{} >{\raggedright\arraybackslash}p{1.7cm} >{\centering\arraybackslash}p{1.85cm} >{\centering\arraybackslash}p{2.0cm} >{\centering\arraybackslash}p{1.85cm} @{}}
		\toprule
		Method & \#Params & Training Time $\downarrow$ & CPSNR $\uparrow$ \\
		\midrule
		w/o VM       & $756.04$ M & $68.49$ min & $4.42$ dB \\
		w/ VM (Ours) & $\mathbf{17.37}$ M & $\mathbf{10.19}$ min & $\mathbf{17.46}$ dB \\
		\bottomrule
	\end{tabular}
	\vspace{-1pt}
\end{table}

\subsection{Ablation Studies}
In this subsection, we evaluate the effectiveness of key design components in Voxel-CKM through ablation studies. All  experiments are conducted in the conference room scenario with $100$ training samples.

\textit{1) Ablation on the VM Decomposition:}
Table \ref{tab:vm-ablation} investigates the effectiveness of the VM decomposition. The variant without VM directly parameterizes the voxelized RF radiance field using dense voxel grids, whereas our Voxel-CKM employs VM decomposition for a low-rank factorized representation. For a fair comparison, both methods use the same voxel grid resolution, with $I=J=K=300$.
As shown in Table \ref{tab:vm-ablation}, incorporating VM decomposition substantially reduces the model size, decreasing the number of learnable parameters by more than an order of magnitude. This compact representation leads to a remarkable improvement in training efficiency, enabling faster convergence during optimization.
More importantly, this improvement does not come at the cost of performance, with a CPSNR of $17.46$ dB achieved.
In contrast, the variant without VM  yields an excessively large model, which  is difficult to optimize and  exhibits inferior performance in the few-shot regime.
These observations confirm the advantage of  the proposed VM decomposition, which provides a compact representation for the voxelized RF radiance field, facilitating efficient CKM construction and improving CSI prediction performance.



\textit{2) Ablation on the Transmitter Prior:} To evaluate the contribution  of the transmitter prior, we compare the CPSNR performance with and without  this prior. As shown in  Table \ref{tab:tx-prior-ablation}, the incorporation of the transmitter prior leads to a noticeable improvement in performance, with CPSNR increasing by over $9$ dB.  This enhancement arises from the  strong inductive bias provided by the transmitter prior, which guides the optimization  of Voxel-CKM under sparse measurements, improving  sample efficiency. Consequently, our Voxel-CKM excels in few-shot learning tasks, demonstrating  potential for practical applications.

\begin{figure*}[t]
	\centering 
	\includegraphics[width=1.0\linewidth]{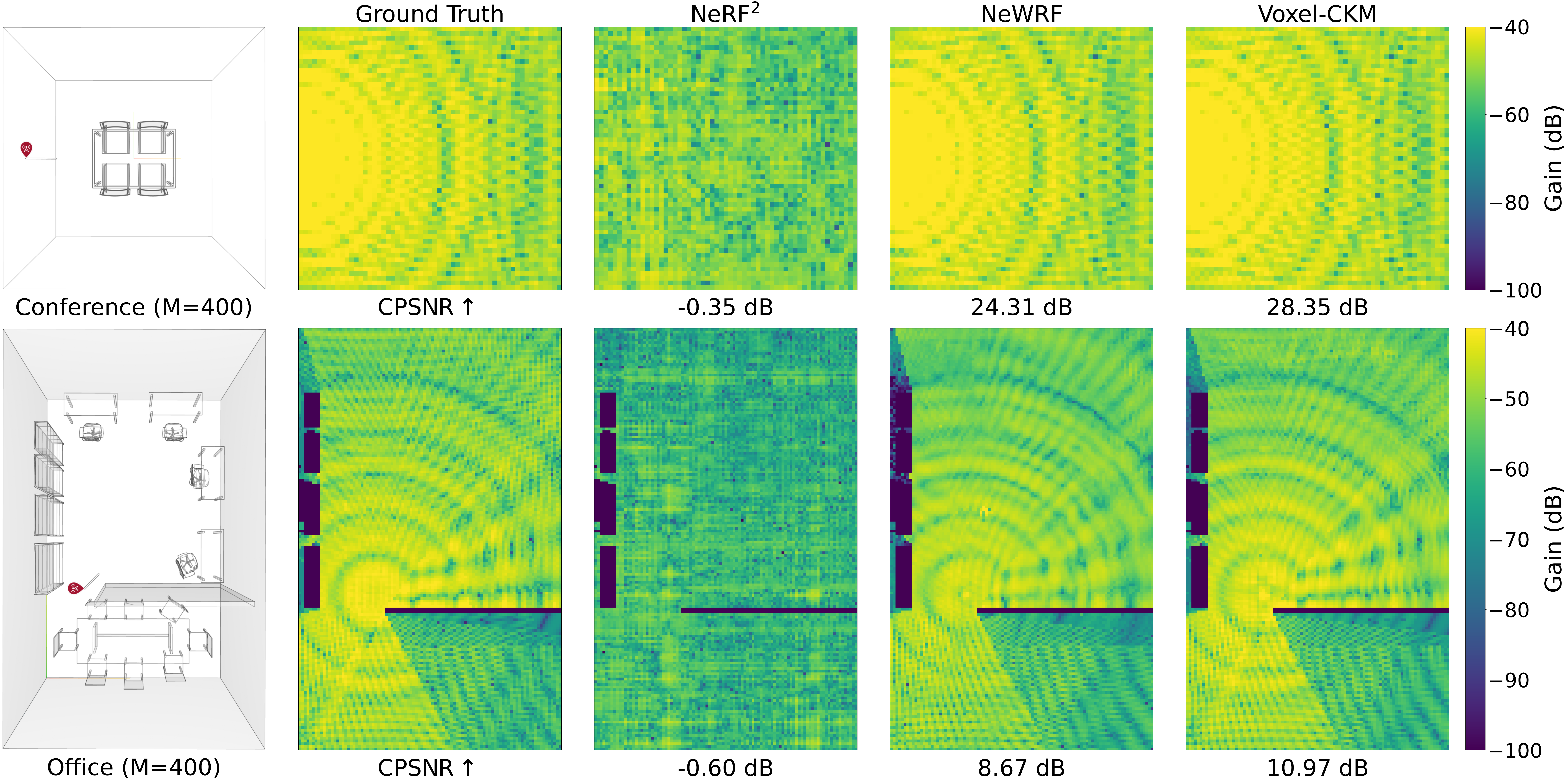} 
	\captionsetup{font={footnotesize}}
	\caption{Visualization of learned CKMs from different models.
		The first column presents the top-down floor plans of two indoor environments.
		The second column shows the corresponding channel gain  heatmaps at a height of $1.2$ m.
		The remaining columns display the learned heatmaps of different models, with each model trained using $400$ samples.
	}
	\label{fig:Visualization} 
	\vspace{0pt}
\end{figure*}

\renewcommand{\arraystretch}{1.5}
\begin{table}[t]
	\centering
	\setlength{\tabcolsep}{1pt}
	\caption{Ablation study on the transmitter prior.}
	\label{tab:tx-prior-ablation}
	\begin{tabular}{
			@{}
			>{\raggedright\arraybackslash}m{3.6cm}
			>{\centering\arraybackslash}m{2.1cm}
			>{\centering\arraybackslash}m{2.1cm}
			@{}
		}
		\toprule
		Method & Training Time $\downarrow$ & CPSNR $\uparrow$ \\
		\midrule
		w/o Transmitter Prior & $\mathbf{10.18}$ min & $8.45$ dB \\
		w/ Transmitter Prior (Ours) & $10.19$ min & $\mathbf{17.46}$ dB \\
		\bottomrule
	\end{tabular}
\end{table}

\renewcommand{\arraystretch}{1.5}
\begin{table}[t]
	\centering
	\caption{Ablation study on the regularized loss function.}
	\label{tab:loss-ablation}
	\vspace{0.5ex}
	\setlength{\tabcolsep}{6pt}
	\begin{tabular}{
			@{}
			>{\raggedright\arraybackslash}m{3.2cm}
			@{\hspace{8pt}}
			>{\centering\arraybackslash}m{2.1cm}
			@{\hspace{8pt}}
			>{\centering\arraybackslash}m{2.1cm}
			@{}
		}
		\toprule
		Loss Function & Training Time $\downarrow$ & CPSNR $\uparrow$ \\
		\midrule
		$\mathcal{L}_{\text{MSE}}$
		& $\mathbf{9.07}$ min
		& $5.72$ dB \\
		$\mathcal{L}_{\text{MSE}}$ + $\mathcal{L}_{\ell_1}$
		& $9.44$ min
		& $6.43$ dB \\
		$\mathcal{L}_{\text{MSE}}$ + $\mathcal{L}_{\text{TV}}$
		& $9.98$ min
		& $16.35$ dB \\
		$\mathcal{L}_{\text{MSE}}$ + $\mathcal{L}_{\ell_1}$ + $\mathcal{L}_{\text{TV}}$ (Ours)
		& $10.19$ min
		& $\mathbf{17.46}$ dB \\
		\bottomrule
	\end{tabular}
\end{table}

\textit{3) Ablation on the Regularized Loss Function:} 
Table \ref{tab:loss-ablation} presents an ablation study on the effect of different loss components in Voxel-CKM.
We observe that Voxel-CKM with the complete loss function outperforms all other configurations. 
When only the MSE loss is used, the model demonstrates reduced stability and inferior performance.
The inclusion  of $\ell_1$ regularization improves generalization by constraining excessive parameter magnitudes in the VM factors, leading to more stable training and a slight increase in CPSNR.
Incorporating TV regularization further enhances the model's performance, contributing to its ability to  penalize large variations between neighboring factors, thus mitigating overfitting.
Together, these loss components substantially improve the robustness and generalization of the model in the few-shot regime.

\subsection{Visualization of Learned CKM}  
Fig. \ref{fig:Visualization} provides qualitative visualizations of the learned CKMs in the conference room and office environments. Specifically, we present the channel gain heatmaps at a height of $1.2$ m for each environment, comparing the ground truth with the predictions of different models.
As shown in  Fig. \ref{fig:Visualization}, Voxel-CKM effectively captures fine-grained  characteristics of  wireless channels, including  interference patterns and coherent spatial variations. Moreover, the learned maps preserve  shadowing and occlusion effects induced by walls, leading to spatially structured channel distributions that align well with the ground truth. These visualizations complement the  quantitative analysis, highlighting the effectiveness of Voxel-CKM in capturing environment-aware representations.

\section{Conclusion}
In this paper, we  proposed Voxel-CKM, a novel framework for fast and few-shot CKM construction.
Specifically, we first formulated a voxelized RF radiance  field that replaces implicit neural representations with explicit voxel grids 
to efficiently capture the spatial variation of wireless channels.
Building upon this, we introduced a compact VM decomposition to parameterize the voxel grids using a small set of matrices and vectors, significantly accelerating the learning process.
To enable   few-shot learning, we  further incorporated a transmitter prior that provides an inductive bias to guide model training under sparse measurements. Additionally,  a regularized loss function was designed to mitigate overfitting and stabilize optimization.
Experimental results demonstrated that the proposed Voxel-CKM achieves substantially faster convergence than existing baselines while delivering superior performance in few-shot scenarios. These results highlight the potential of Voxel-CKM as an efficient and practical solution for CKM construction in future wireless communication systems.

\bibliographystyle{IEEEtran}           
\bibliography{IEEEabrv,Reference}      

\end{document}